\documentclass[review]{elsarticle}
\usepackage[a4paper,left=2.5cm,right=2.5cm,top=2.5cm,bottom=2.5cm]{geometry}
\usepackage[utf8]{inputenc}
\usepackage[T1]{fontenc}
\usepackage[mathcal]{eucal}
\usepackage{amsfonts, amsmath, amssymb, bm, mathrsfs}
\usepackage{graphicx, epsfig, color, colortbl, subfigure, adjustbox}
\usepackage{array, import, float, multirow}
\usepackage{relsize, lscape, rotating}
\usepackage{natbib}                   
\usepackage{ulem, soul, lmodern}
\usepackage[hidelinks]{hyperref}

\hypersetup{ pdftitle  = {LDPM modeling of pressure-dependent inelasticity in granular rocks},
             pdfauthor = {S. Esna Ashari} }

\biboptions{sort&compress}	
\journal{International Journal of Rock Mechanics and Mining Sciences}

\begin{document}

\begin{titlepage}
\clearpage\thispagestyle{empty}
\noindent
\hrulefill
\begin{figure}[h!]
\centering
\includegraphics[width=2 in]{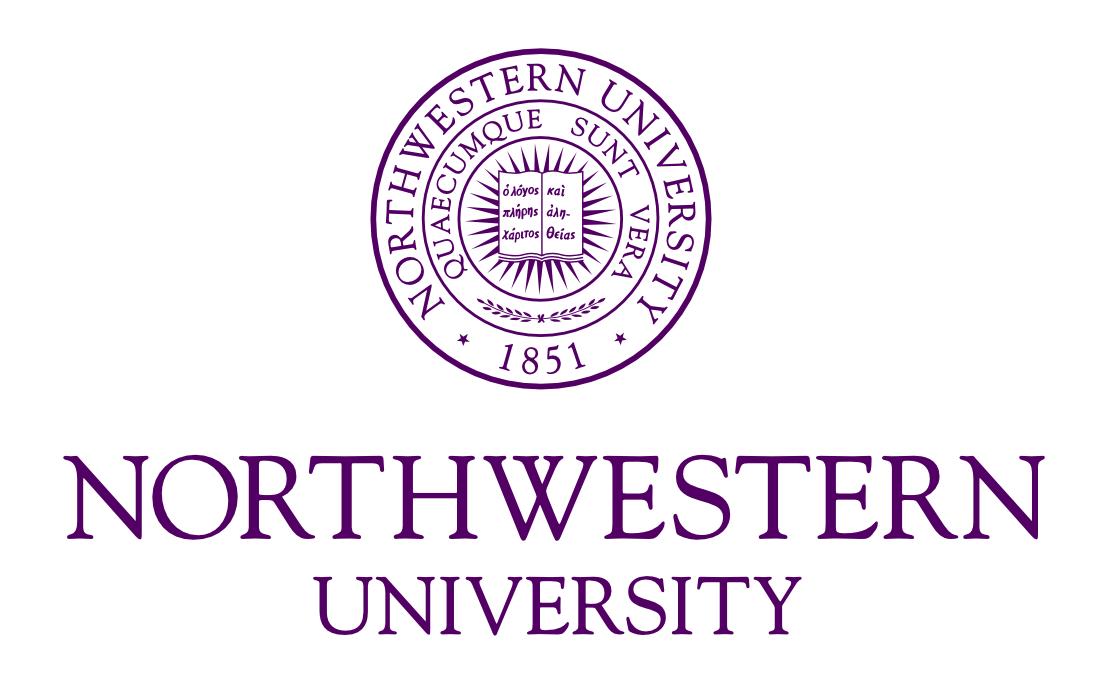}
\end{figure}
\begin{center}
{
{\bf Center for Sustainable Engineering of Geological and
Infrastructure Materials (SEGIM)} \\ [0.1in]
Department of Civil and Environmental Engineering \\ [0.1in]
McCormick School of Engineering and Applied Science \\ [0.1in]
Evanston, Illinois 60208, USA
}
\end{center} 
\hrulefill \\ \vskip 2mm
\vskip 0.5in
\begin{center}
{\large {\bf \uppercase{Lattice Discrete Particle Model (LDPM) for pressure-dependent inelasticity in granular rocks}}}\\[0.5in]
{\large {\sc Shiva Esna Ashari, Giuseppe Buscarnera, Gianluca Cusatis}}\\[0.75in]
{\sf \bf SEGIM INTERNAL REPORT No. 16-02/478L}\\[0.75in]
\end{center}
\vskip 1.2in
\noindent {\footnotesize {{ Submitted to International Journal of Rock Mechanics and Mining Sciences \hfill February 2016} }}

\end{titlepage}

\newpage
\clearpage \pagestyle{plain} \setcounter{page}{1}
\begin{frontmatter}
\title{\textbf{Lattice Discrete Particle Model (LDPM) for pressure-dependent inelasticity in granular rocks}}
\author[NWU]{S. Esna Ashari}
\ead{s-esnaashari@u.northwestern.edu}
\author[NWU]{G. Buscarnera\corref{correspondingauthor}}
\cortext[correspondingauthor]{Corresponding author}
\ead{g-buscarnera@northwestern.edu}
\author[NWU]{G. Cusatis}
\ead{g-cusatis@northwestern.edu}
\address[NWU]{Northwestern University, Department of Civil and Environmental Engineering, USA}
%
%
%
%
%
\begin{abstract}
This paper deals with the formulation, calibration, and validation of a Lattice Discrete Particle Model (LDPM) for the simulation of the pressure-dependent inelastic response of granular rocks. LDPM is formulated in the framework of discrete mechanics and it simulates the heterogeneous deformation of cemented granular systems by means of discrete compatibility/equilibrium equations defined at the grain scale. A numerical strategy is proposed to generate a realistic microstructure based on the actual grain size distribution of a sandstone and the capabilities of the method are illustrated with reference to the particular case of Bleurswiller sandstone, i.e. a granular rock that has been extensively studied at the laboratory scale. LDPM micromechanical parameters are calibrated based on evidences from triaxial experiments, such as hydrostatic compression, brittle failure at low confinement and plastic behavior at high confinement. Results show that LDPM allows exploring the effect of fine-scale heterogeneity on the inelastic response of rock cores, achieving excellent quantitative performance across a wide range of stress conditions. In addition, LDPM simulations demonstrate its capability of capturing different modes of strain localization within a unified mechanical framework, which makes this approach applicable for a wide variety of geomechanical settings. Such promising performance suggests that LDPM may constitute a viable alternative to existing discrete numerical methods for granular rocks, as well as a versatile tool for the interpretation of their complex deformation/failure patterns and for the development of continuum models capturing the effect of micro-scale heterogeneity.
\end{abstract}
\end{frontmatter}
%
%
%
\section{Introduction}
Granular rocks display complex mechanical properties,  such as the transition from brittle to ductile response upon increasing confinement \cite{wong_2012}, the tendency to dilate or contract  upon shearing \cite{menendez_1996, wong_1997}, and the formation of a wide range of strain localization mechanisms \cite{baud_2004, wong_2001, fossen_2007}.  Such rich variety of deformation modes depends on the inelastic properties of rocks, and it is invariably controlled by the confining pressure. For example, while localized dilatant faulting is typically observed at low confinements, delocalized shear-enhanced compaction often characterizes the deformation response at high pressures. The transition from one type of response to another is typically gradual \cite{wong_2012} and poses considerable challenges due to the competition of the microscopic processes that characterize each of the two aforementioned macroscopic phenomena. This intermediate behavior has been found to be crucial for a variety of applications, including the tectonics of faulting \cite{Aydin_1978, Jamison_1982, Shipton_2001}, the coupling between strain localization and fluid flow \cite{Knipe_1997, Wong_1999}, reservoir compaction \cite{Bouteca_1996, nagel_2001} and borehole instability \cite{Veeken_1989, Coelho_2005}. \\
In the brittle faulting regime, the onset of dilatancy is associated with the propagation of cracks that align along directions subparallel to the maximum compressive stress. The coalescence of these cracks, as well as the frictional interaction between fractured and unfractured zones, ultimately lead to the onset of  persistent shear bands, as well as to changes in physical properties, such as stiffness, permeability, and electrical conductivity \cite{Dresen_2004, Paterson_2005}. In the cataclastic flow regime, grain crushing and pore collapse dominates the deformation process, ultimately leading to extensive densification of the rock mass. 
In high-porosity rocks, such micro-mechanical processes have been found to promote compaction bands, i.e.  modes of strain localization characterized by the accumulation of compressive strains into narrow zones \cite{Mollema_1996, olsson_1999}. While these compaction localization processes are induced by a local loss of strength, the rearrangement of crushed fragments and the reduction of the local porosity often lead to a gradual transition to a delocalized mode of deformation \cite{das_2014}. As a result, unlike single shear bands, multiple compaction zones may propagate across the sample until a complete re-hardening of the specimen is observed \cite{holcomb_2003}.\\
The prevalence of a specific form of microscopic damage depends on the microstructural attributes of a rock (e.g., grain size and sorting porosity; degree of cementation), as well as by its inherent heterogeneity. Discrete mechanical methods are therefore convenient tools to accommodate grain-scale attributes and explain their impact on the macroscopic deformation of rock cores. For example, the discrete element method (DEM) has proved to be an effective tool for simulating the micromechanics of unconsolidated materials, such as soil, sediment and fault gouge \cite{Cundall_1979, Antonellini_1995, Morgan_1999, Aharonov_2002}. DEM represents the material as an assemblage of independent particles interacting through forces computed on the basis of frictional contact models.
Such methods have often been adapted to the case of lithified geomaterials by incorporating inter-particle bonds accounting for the presence of cementation, thus mimicking the nucleation of cracks though the brittle failure of cohesive cement bridges \cite{Potyondy_1996, Potyondy_2004}. Such enhancements have enabled DEM to simulate complex processes such as the development of shear bands and brittle fracturing \cite{Schopfer_2013, Scholtes_2013}.\\
Nevertheless, standard DEM  techniques based on spherical particles tend to produce unrealistically ratios of uniaxial compressive to tensile strength \cite{Altindag_2010}, thus hampering the satisfactory prediction of the failure characteristics of granular rocks deformed in the brittle faulting regime. Although this problem can be mitigated by increasing the density of bonds between particles \cite{Donze_1997, Scholtes_2013} or by magnifying the grain interlocking  through irregularly shaped particles \cite{Cho_2007, Lan_2010}, the ability to capture the full spectrum of tensile and/or compressive failure mechanisms through a unified framework still represents a major challenge. Similar limitations exist also for simulations in the high-pressure regime, where DEM analyses are often used in conjunction with computationally intensive particle replacement schemes mimicking the effect of grain crushing  \cite{McDOWELL_2013}. While these approaches have provided insights into the interpretation of compaction localization, they often involve an unrealistic loss of grain mass, thus preventing a realistic simulation of crushing-induced hardening upon hydrostatic compression \cite{wong_2012}.\\
To tackle these problems, this paper proposes an alternative discrete method that, by relying on a direct representation of the microstructure, aims to accommodate a wide range of inelastic mechanisms i.e., it enables accounting simultaneously for brittle/dilative modes of failure, as well as for the plastic regime of compactive deformation. The proposed approach builds upon the so-called Lattice Discrete Particle Model (LDPM), successfully developed by Cusatis and coworkers \cite{Cusatis_2011_1, Cusatis_2011_2} for the simulation of failure processes in quasi-brittle solids such as concrete. A noticeable feature of LDPM is its ability to simulate a granular microstructure through a system of polyhedral particles connected through a three-dimensional lattice. Such particles can be placed randomly across the volume in accordance with a prescribed grain size distribution, thus enabling the direct representation of a heterogeneous system of grains surrounded by a bonding agent (e.g., mortar in concrete or mineral precipitants in natural rocks). At variance with DEM techniques, the kinematics of the skeleton is modeled on the basis of the displacements and rotations computed at the nodes of the lattice, thus enabling the computation of strain components oriented normally and/or tangentially to the facets between the polyhedral particles. Such hypotheses imply the use of an internal kinematics substantially different from that of DEM. This facilitates the use of more sophisticated constitutive laws to model the forces transferred among adjacent particles. Recent works have demonstrated the ability of this approach to reproduce various aspects of quasi-brittle behavior, such as fracture initiation and propagation, shear banding, and frictional processes \cite{Alnaggar_2013, Smith_2014, Shiva_2015, Rezakhani_2016}. Therefore, LDPM offers a convenient platform to simulate the mechanics of sandstones, a particular class of quasi-brittle solids for which the pressure-dependent inelastic properties are primarily controlled by the heterogeneity of their grain skeleton. Although the strategy discussed hereinafter is in principle applicable to the analysis of any type of granular rock, here its capabilities are discussed for the particular case of Bleurswiller sandstone, that is a high-porosity rock extensively studied in the literature and for which a wide range of strain localization mechanisms have been documented \cite{fortin_2005, fortin_2009}.

%
\section{Grain generation}\label{LDPM_graingeneration}
\begin{figure}[!ht]
\centering
\includegraphics[width=16.0cm]{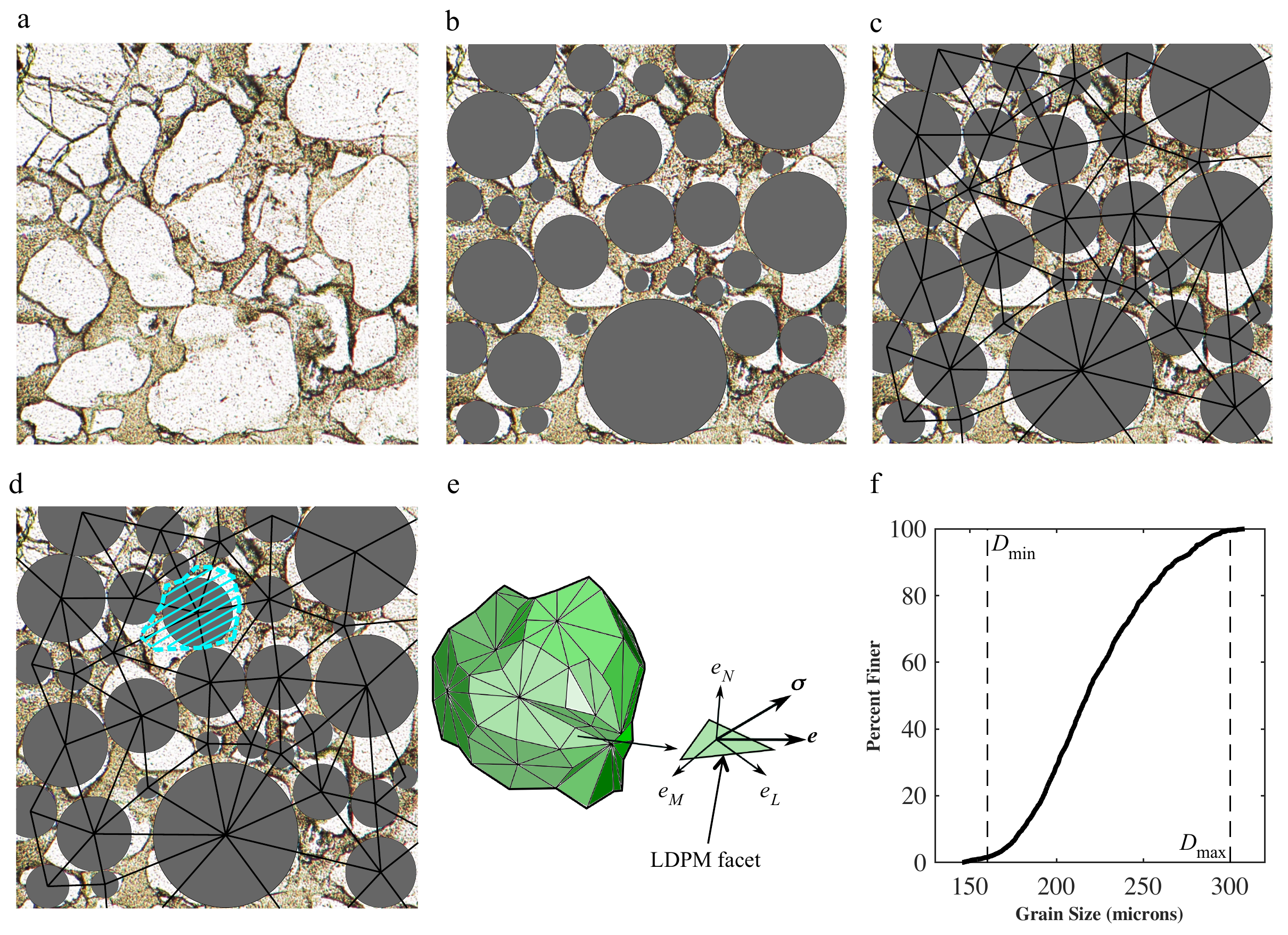}
\caption{(a) Microstructure of a sandstone. (b) Artificial supporting system for grain generation. (c) 2D representation of Delaunay tetrahedralization for the supporting system. (d) 2D representation of a polyhedral cell. (e) A 3D polyhedral cell/a LDPM cement-coated grain. (f) Grain size distribution of Bleurswiller sandstone used for the LDPM simulations.}\label{figure1}
\end{figure}

The strategy adopted in LDPM to replicate the grain-scale heterogeneity of sandstones is schematically depicted in Figure 1, which illustrates through a simplified two-dimensional representation the basic steps required to map the real microstructure of a granular rock into its numerical analogue.\\
 As the grains in sandstones tend to be closely packed and in direct contact with each other (Figure \ref{figure1}a), the isolation of cement bridges and grains is not straightforward. Therefore, a reasonable simplification to discretize the domain into cement-coated grains having the same size distribution of the actual grains can be obtained by hypothesizing that the granular lattice controlling the micro-mechanical interactions is only secondarily affected by the geometry of the cement bridges. From a modeling standpoint, this choice implies that the contribution of the cementing phase will not be modeled explicitly, but it will rather be embedded implicitly into the particle-scale constitutive laws controlling the interaction between skeletal grains.\\
In LDPM, the geometrical characterization of the mesostructure of sandstones is constructed by means of an artificial supporting system based on spherical particles placed at the center of sandstone grains (Figure \ref{figure1}b). Such a supporting system is generated by following a strategy similar to that proposed by  \citet{Cusatis_2011_1} for the case of concrete, i.e. by defining the size distribution of the spherical supports through a probability density function (pdf) defined as follows:
\begin{equation}\label{pdf}
{f}(d) = \frac{qd_0^q}{[1-(d_0/d_a)^q]d^{q+1}}
\end{equation}
which is associated with a sieve curve in the form:
\begin{equation}\label{pdf}
{F}(d) = \Big(\frac{d}{d_a}\Big)^{n_{F}}
\end{equation}
where $d_0$ is the minimum particle size, $d_a$ is the maximum particle size, $q$ is a material parameter and $n_{F}=3-q$ is the sieve curve exponent. The volume fraction of simulated particles ($\nu_{a0}$) can be calculated as:
\begin{equation}
\nu_{a0} = [1-F(d_0)]\nu_a=[1-\Big(\frac{d_0}{d_a}\Big)^{n_{F}}]\nu_a
\end{equation}
where $\nu_a$ is the particle volume fraction per unit volume of sandstone and the total volume of simulated particles is $V_{a0}=\nu_{a0}V$, if $V$ is the volume of the domain of interest. It should be noted that for each sandstone, $d_0$, $d_a$, $n_{F}$ and $\nu_a$ should be calibrated based on measured grain size distribution of the rock.\\
In this approach, the particle diameters $d_i$ are computed by sampling the cumulative distribution function (cdf) associated with Equation (\ref{pdf}) by means of a random number generator \cite{Cusatis_2011_1}.  New particles are generated until the total volume of generated spherical particles $\widetilde{V}_{a0}=\sum_i(\pi d_i^3/6)$, exceeds $V_{a0}$. \\
After the stage of particle generation, the particles are randomly distributed across the specimen on vertices, edges, surface faces, and interior volume. In order to have a statistically isotropic random mesostructure, particle centers are placed throughout the volume of the specimen from the largest to the smallest, preventing possible overlaps between the particles.\\
The next step is to define the topology of the grains of the modeled sandstone by using Delaunay tetrahedralization and a 3D tessellation. Through the Delaunay tetrahedralization, the nodal coordinates of the particle centers are used to define a three-dimensional mesh of tetrahedra (Figure \ref{figure1}c). These tetrahedra do not overlap, fill the entire volume of the specimen, and have vertices coinciding with the given particle centers. The final geometry of the grains is defined by performing a 3D tessellation of the domain anchored to the Delaunay tetrahedralization. For details on the adapted tessellation, the reader is refereed to Ref. \cite{Cusatis_2011_1}. By collecting all the facets associated with one particle (Figure \ref{figure1}d), it is possible to obtain a polyhedral cell representing a cement-coated grain (Figure \ref{figure1}e). The grain size distribution of the simulated sandstone can eventually be expressed by computing the volume of each polyhedral cell and plotting the statistical distribution of their volume-equivalent sphere diameters similar to Figure \ref{figure1}f. It is worth noting that the iterative comparison between this synthetic grain size distribution and the actual grading of the rock is pivotal to define the parameters ($d_0$, $d_a$, $n_{F}$, $\nu_a$), which must in turn be calibrated through a trial-and-error procedure specific for the selected rock. \\

\section{LDPM constitutive equations}\label{LDPM_constitutive}
In LDPM, the grains interact with each other through the facets that connect them, and the displacement field is defined through the rigid body kinematics of the grains. Similar to previous LDPM work \cite{Cusatis_2011_1}, the mechanics of grain interaction is formulated based on an analysis of an assemblage of four particles located at the vertices of a tetrahedron. The displacements and rotations of the nodes adjacent to a facet can be used to compute the displacement jump $[\![\mathbf{u}_{c}]\!]$ at the centroid of each facet in the tetrahedron. Such displacement jump is then used to define the strain components of the facet (Figure \ref{figure1}e):
\begin{equation}
e_{N} = \frac{\mathbf{n}^{T}[\![\mathbf{u}_{c}]\!]}{\ell}; \;e_{L} = \frac{\mathbf{l}^{T}[\![\mathbf{u}_{c}]\!]}{\ell}; \;e_{M} = \frac{\mathbf{m}^{T}[\![\mathbf{u}_{c}]\!]}{\ell}
\end{equation}
where $\ell$ indicates the interparticle distance, $\mathbf{n}$, $\mathbf{l}$ and $\mathbf{m}$ are unit vectors that define a local reference system attached to each facet. Note that the displacement jump $[\![\mathbf{u}_{c}]\!]$ is defined such that positive normal strain $e_N$ represents compression.\\
Prior to the initiation of micro-scale inelastic processes, the constitutive relation between the strain vector $\bm{e}$ and the stress vector $\bm{\sigma}$ at the facet level is incrementally elastic:
\begin{equation}\label{elastic}
\dot{\sigma}_{N} = E_N \dot{e}_N;\;\dot{\sigma}_{L} = \alpha E_N \dot{e}_L;\;\dot{\sigma}_{M} = \alpha E_N \dot{e}_M
\end{equation}
where $E_N$ is the effective normal modulus, $\alpha$ is the shear-normal coupling parameter.\\
In LDPM, the reversible elastic behavior is limited by a number of nonlinear stress-strain boundaries, each mimicking different types of meso-scale inelastic phenomena that involve softening for pure tension and shear-tension, as well as plastic hardening for pure compression and shear compression. 
\subsection{Pore collapse and material compaction}
Under high-pressure hydrostatic compression, sandstones exhibit strain-hardening plasticity, which is characterized by an initial phase of pores collapse and a later phase, in which the walls of completely collapsed pores become in contact leading to a significant densification of the material. In terms of stress strain response, the first phase is associated with a sudden decrease of the stiffness yielding that is later regained in the second phase (rehardening). LDPM simulates these phenomena through a strain-dependent normal boundary ($\sigma_{bc}$) limiting the compressive normal stress and it is assumed to be a function of the local volumetric strain $e_V$ and deviatoric strain $e_D$. The volumetric strain is computed at the tetrahedron level as $e_V = (V-V_0)/V_0$, where $V$ and $V_0$ are the current and initial volume of the tetrahedron, respectively. In each LDPM tetrahedron, all twelve facets are assumed to be subjected to the same volumetric strain, whereas each facet is characterized by a different value of the deviatoric strain calculated by subtracting the volumetric strain from the normal strain: $e_D = e_N-e_V$. This definition of local strains are equivalent to the ones used in typical microplane model formulations \cite{bazant_2000, bazant_2013}\\
For a constant deviatoric-to-volumetric strain ratio, $r_{DV} = e_D/e_V$, the pre-yielding response, is assumed to be characterized by an initial bilinear evolution modeling the closure of existing fissures. This stage is followed by linear elastic response after the complete closure of the fissures (Figure \ref{figure2}a). The post-yield response is assumed to be controlled by a linear plastic behavior modeling the initial stages of pore collapse, then switching to an exponential form to model compaction-induced rehardening (Figure \ref{figure2}a).  The relations that simulate such sequence of compression processes is:
\begin{equation}
\sigma_{bc}\left(e_D,e_V\right)=\left\lbrace
\begin{matrix}
& \max(E_N e_N, \beta_1 E_N e_N) & 0 \leq e_V \leq e_{c0}+e_f\\ 
& \sigma_{c0}+\langle e_N-(e_{c0}+e_f) \rangle H_c\left(e_V,e_D\right) & e_{c0}+e_f \leq e_N \leq e_{c1}+e_f\\
& \sigma_{c1} \left(r_{DV}\right) \exp [ \left(e_N-(e_{c1}+e_f)\right) \beta_2 H_c\left(e_V,e_D\right)/\sigma_{c1} \left(r_{DV}\right)] & \text{otherwise}
\end{matrix}
\right.
\end{equation}
where $\beta_1$ is the fissure closure parameter used to enforce the initial nonlinearity and $e_f$ is the normal strain offset associated with the fissure closures at which the typical linear elastic response commences.  In the present model, $e_f$ is assumed to be a function of $\ell$, the local interparticle distance, and $w_0$, the average size of fissure cracks openings in a sandstone: $e_f=w_0/\ell$. The parameter $\sigma_{c0}$ is the meso-scale yielding stress at the onset of pore collapse and $e_{c0}+e_f$ is the corresponding compaction strain; $H_c\left(e_V,e_D\right)$ is the initial hardening modulus, $\beta_2$ the rehardening coefficient, $e_{c1}+e_f$ the compaction strain at which rehardening begins with $\sigma_{c1} \left(r_{DV}\right)=\sigma_{c0}+(e_{c1}-e_{c0})H_c\left(e_V,e_D\right)$ as the correlated stress. The hardening modulus is formulated through Equation \ref{H_c} which preserves the continuity of the slope for transition from positive to negative deviatoric-to-volumetric strain ratio and vice versa \cite{Ceccato_2015} and enables the model to simulate the observed post-yield horizontal plateau featured by typical experimental data relevant to triaxial tests:
\begin{equation}\label{H_c}
H_c\left(e_V,e_D \right)=\left\lbrace
\begin{matrix}
& \frac{H_{c0}-H_{c1}}{1+\kappa_{c2}\langle r_{DV1}-\kappa_{c1}\rangle}+H_{c1} & e_V \geq0 &  \text{(contraction)}\\ 
& \frac{H_{c0}-H_{c1}}{1+\kappa_{c2}\langle r_{DV2}-\kappa_{c1}\rangle}+H_{c1} & e_V <0 &  \text{(expansion)}\\
\end{matrix}
\right.
\end{equation}
where $H_{c0}$ is a material parameter, $\kappa_{c1}=1$ and $\kappa_{c2}=5$ \cite{Cusatis_2011_1} and
\begin{equation}
r_{DV1}=-\frac{|e_D|}{e_V-e_{V0}};\;  r_{DV2}=-\frac{|e_D|}{e_{V0}}
\end{equation}
with $e_{V0}=\kappa_{c3}e_{c0}$ and $H_{c1}=\kappa_{c3}E_N$ \cite{Ceccato_2015}. For the sake of simplicity, in this paper, $\kappa_{c3}$ is assumed to be zero.  As previously mentioned, it must be noted that the meso-scale constitutive relations listed above encapsulate the variety of fine-scale processes that take place at the interface between grains and/or within each single cement-coated particle. As a result, their parameters must be considered as an outcome of the constitution of the sandstone grains, thus reflecting indirectly the role of sub-resolution parameters that are not explicitly modeled by the LDPM (e.g., cement porosity, intra-grain cracks, etcetera).
\subsection{Frictional behavior}
Frictional phenomena can be simulated effectively through classical incremental plasticity. The incremental shear stresses are computed as
\begin{equation}
\dot{\sigma}_L = \alpha E_N \left(\dot{e}_L-\dot{e}_L^p\right);\;\dot{\sigma}_M = \alpha E_N \left(\dot{e}_M-\dot{e}_M^p\right)
\end{equation}
Tangential plastic strain increments are assumed to obey the normality rule $\dot{e}_L^p=\dot{\lambda}\partial\varphi/\partial\sigma_L;\; \dot{e}_M^p=\dot{\lambda}\partial\varphi/\partial\sigma_M$, where $\lambda$ is the plastic multiplier. An independent plastic flow as described in the previous section is assumed to be active along the direction normal to the facets, thus implying the lack of normality in terms of normal plastic strains. This hypothesis implies that the macroscopic plastic dilatancy is not directly enforced at the meso-scale, but it is rather simulated as an emerging attribute linked to the degree of grain interlocking of the numerical lattice. The plastic potential is defined as $\varphi=\sqrt{\sigma_L^2+\sigma_M^2}-\sigma_{bs}\left(\sigma_N\right)$, where the nonlinear frictional law for the shear yielding stress is assumed to be
\begin{equation}\label{sigma_bs}
\sigma_{bs}= \sigma_{s}-\left(\mu_0-\mu_\infty \right)\sigma_{N0}+\mu_\infty\sigma_{N}+\left(\mu_0-\mu_\infty \right)\sigma_{N}\exp \left(\sigma_{N}/\sigma_{N0}\right)
\end{equation}
In Equation \ref{sigma_bs}, $\sigma_{s}$ is the cohesion, $\mu_0$ and $\mu_\infty$ are the initial and final internal friction coefficients and $\sigma_{N0}$ is the normal stress at which the internal friction coefficient transitions from  $\mu_0$ to $\mu_\infty$ which basically
governs the nonlinearity of the shear boundary. It can be seen that in the presence of compressive stresses, the shear strength increases due to frictional effects (Figure \ref{figure2}b). It is worth mentioning that the classical linear (Coulomb-type) frictional law with slope $\mu_0$ or $\mu_\infty$ is obtained by setting $\sigma_{N0}=\infty$ or $\sigma_{N0}=0$, respectively. The frictional law is also linear for $\mu_0=\mu_\infty$ for any values of $\sigma_{N0}=0$.
\subsection{Fracturing behavior}
\begin{figure}[!ht]
\centering
\includegraphics[width=16.2cm]{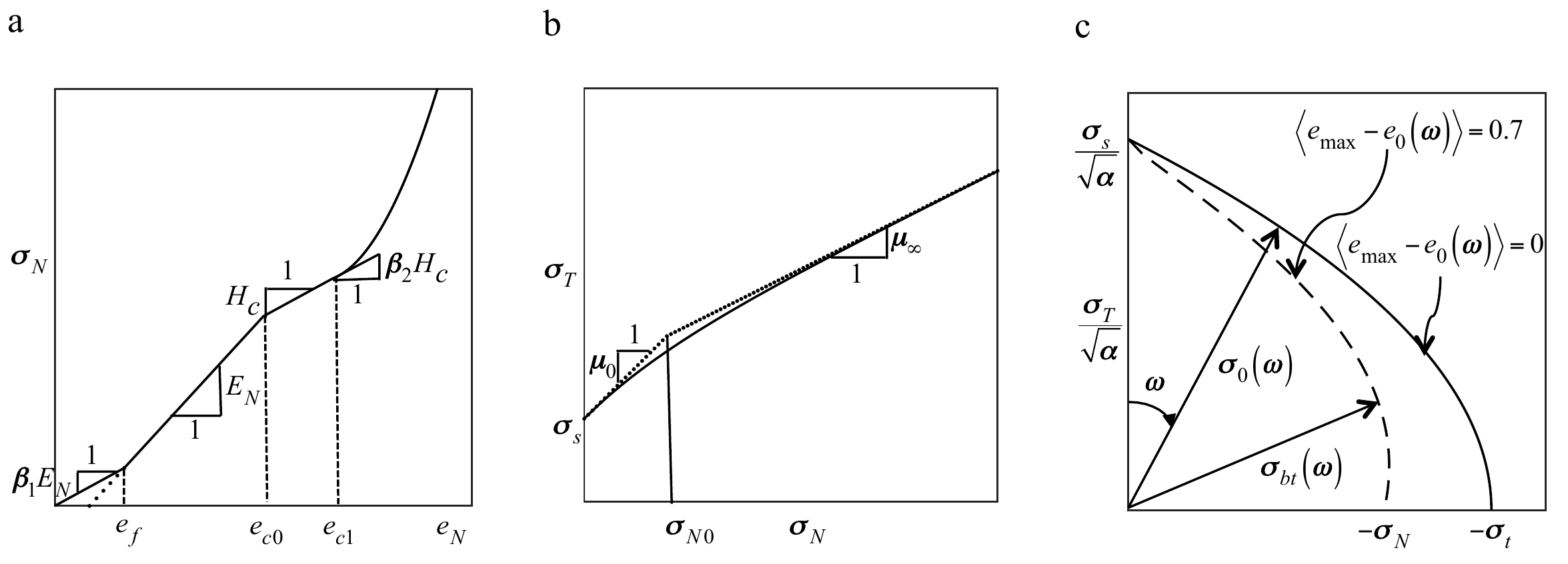}
\caption{(a) Normal stress versus normal strain in compression. (b) Shear strength as a function of normal compressive stresses. (c) Shear strength as a function of normal tensile stresses.}\label{figure2}
\end{figure}

For fracturing behavior characterized by tensile normal strains ($e_N < 0$), the fracture evolution is formulated through the relationship between the effective strain $e$, $e = \sqrt{e_N^2+\alpha \left(e_L^2+e_M^2 \right) }$, and the effective stress $\sigma$, $\sigma = \sqrt{\sigma_N^2+ \left(\sigma_L^2+\sigma_M^2 \right)/\alpha}$, which define the normal and shear stresses as
\begin{equation}
\sigma_{N} = e_N\frac{\sigma}{e};\;\sigma_{L} = \alpha e_L\frac{\sigma}{e};\;\sigma_{M} = \alpha e_M\frac{\sigma}{e}
\end{equation}
The strain-dependent limiting boundary for this type of behavior is formulated through an exponential decay, see Equation (\ref{Fracture_boundary}), and enforced through a vertical (at constant strain) return algorithm. One can write
\begin{equation}
\sigma_{bt}\left(e,\omega \right)= \sigma_{0}\left(\omega\right) \exp \Bigg[-H_0\left(\omega\right)\frac{\langle e_{max}-e_0\left(\omega\right) \rangle }{\sigma_0 \left(\omega\right)}\Bigg] \label{Fracture_boundary}
\end{equation}
where $e_{max}$ is the maximum effective strain attained during the loading history and $\omega$ is the coupling variable that represents the degree of interaction between shear and normal loading, defined as $\tan(\omega)= -e_N / {\sqrt{\alpha} e_T} $ in which $e_T=\sqrt{e_M^2+e_L^2}$ is the total shear strain.
The function $\sigma_{0}\left(\omega\right)$ is the strength limit for the effective stress
\begin{equation}
\sigma_{0}\left(\omega\right) =  \sigma_t \frac{-\sin(\omega)+\sqrt{\sin^2(\omega)+4\alpha \cos^2(\omega)/r^2_{st}}}{2\alpha \cos^2(\omega)/r^2_{st}}\label{yield_boundary}
\end{equation}
where $r_{st}=\sigma_s/\sigma_t$ is the ratio between the shear (cohesion) to tensile strength. Equation (\ref{yield_boundary}) is a parabola in $\sigma_N-\sigma_T$ space with its axis of symmetry along the $\sigma_N$-axis (Figure \ref{figure2}c).
The exponential decay of the boundary $\sigma_{bt}$ starts when the maximum effective strain reaches its elastic limit $e_0\left(\omega\right)=\sigma_{0}\left(\omega\right)/E_N$, and the decay rate is governed by the post-peak slope (softening modulus) $H_0\left(\omega\right)$:
\begin{equation}
H_0\left(\omega\right) =  H_t\left( \frac{2\omega}{\pi}\right)^{n_t}
\end{equation}
$n_t$ is the softening power and $H_t$ is the softening modulus in pure tension ($\omega=\pi/2$) expressed as $H_t =  2E_N/(\ell_t/\ell-1)$ where $\ell_t=2E_NG_t/\sigma^2_t$, $G_t$ is the meso-scale fracture energy, and $\ell$ is the length of the tetrahedron edge (or interparticle distance) associated with the facet of interest.

\section{Calibration and validation}\label{LDPM_calibration}
In this section, the constitutive parameters of the LDPM are calibrated and validated for Bleurswiller sandstone. The experimental data used for calibration and validation purposes are derived from previous studies on cylindrical specimens of this rock (one for each test) with the diameter of 40 mm and the height of 80 mm, reported in \citet{fortin_2005} and \citet{fortin_2009}. All the simulations are done on a cylindrical specimen with the diameter and height of 2.4 mm as the representative volume with 3,400 simulated grains. The simulation of the actual sample size would lead to over 18,000,000 grains and to excessive computational cost. Reported data about the grain size distribution of this sandstone indicate $D_{max}=300$ \text{$\mu$m} and $D_{min}=160$ \text{$\mu$m} and the mean diameter of 220 \text{$\mu$m}  \cite{fortin_2005}. As a result, the algorithm for the generation of the LDPM granular lattice discussed in Section \ref{LDPM_graingeneration} was used to approximate these limits. The outcome of the trial-and-error calibration procedure with $d_0$=90 \text{$\mu$m}, $d_a$=230 \text{$\mu$m}, $n_{F}$=0.5, $\nu_a$=0.58 generates a grain size distribution with a minimum diameter of 145 \text{$\mu$m}, a maximum diameter of 305 \text{$\mu$m}, and a mean diameter of 220 \text{$\mu$m} (Figure \ref{figure1}f). 

\subsection{Hydrostatic test}
\begin{figure}[!ht]
\centering
\includegraphics[width=12.2cm]{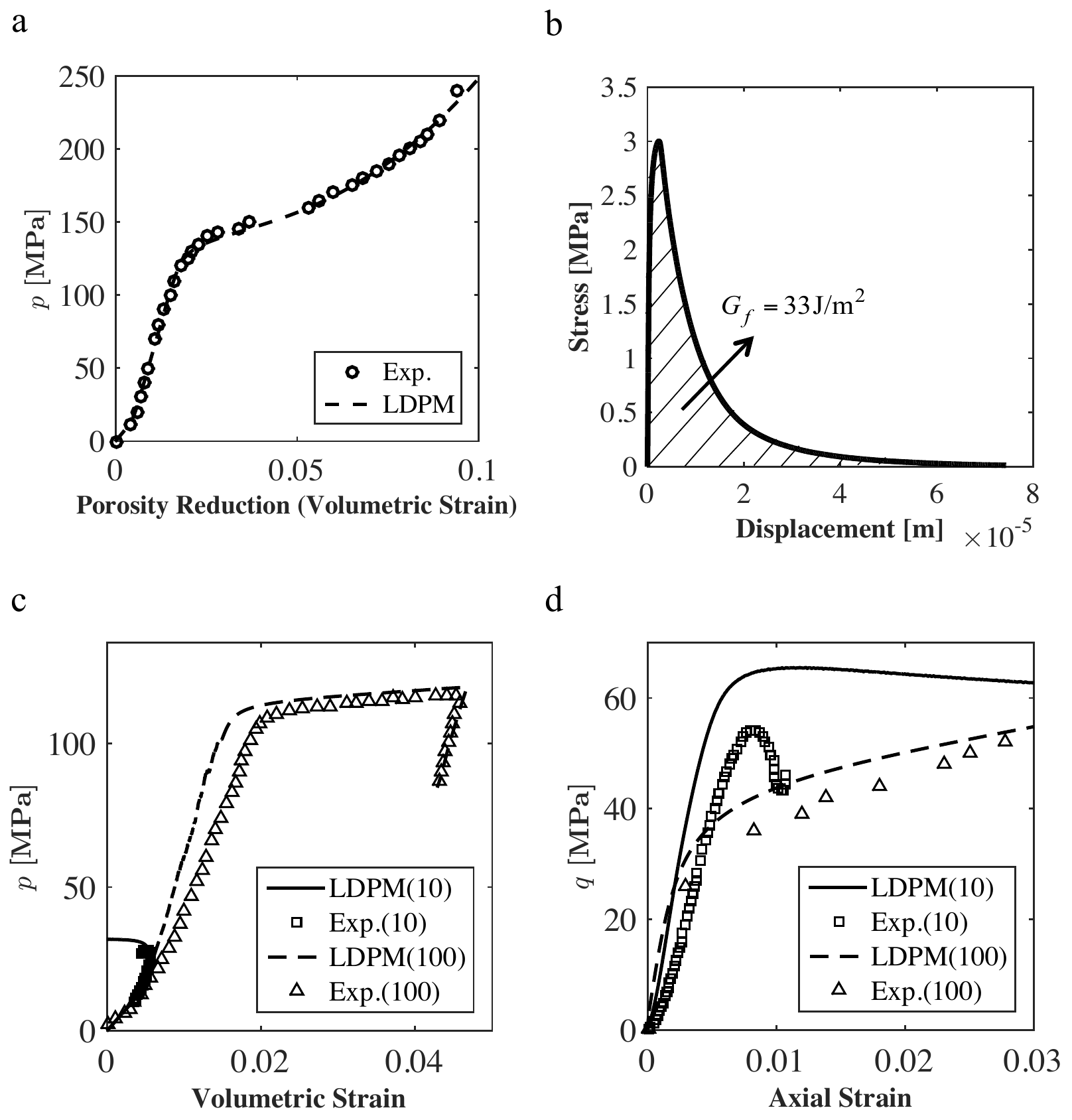}
\caption{(a) Experimental data (after \cite{fortin_2005}) and LDPM simulation of hydrostatic test. (b) Load-displacement response of the direct tension test. (c) Mean stress vs volumetric strain for triaxial tests with 10 and 100 MPa confinement pressures. (d) Deviatoric stress vs axial strain for triaxial tests with 10 and 100 MPa confinement pressures.}\label{figure3}
\end{figure}

The first step of the calibration process involves the parameters governing compression on the LDPM facets. A hydrostatic test is therefore considered, with the goal to restrain the analysis to a stress path mobilizing prevalently volumetric compressive loading along the normal LDPM facets. The response measured from a hydrostatic test on Bleurswiller sandstone is used to calibrate the parameters that control the elastic and compressive response; the normal modulus ($E_N$=27,155 MPa), the fissure closure parameter ($\beta_1$=0.34), the average size of fissure crack opening ($w_0$=0.2 \text{$\mu$m}), the yielding compressive stress ($\sigma_{c0}$=148 MPa), the rehardening coefficient ($\beta_2$=3.5), the initial hardening modulus ($H_{c0}$=2,037 MPa) and $e_{c1}$=4.5$e_{c0}$. In addition, the shear-normal coupling parameter ($\alpha$=0.167) is identified by the value of Poisson's ratio ($\nu$=0.2) calculated from shear modulus ($G$=4,000 MPa) and bulk modulus ($K$= 5,000 MPa) \cite{fortin_2005}. The simulated response for the hydrostatic test on Bleurswiller sandstone is reported in Figure \ref{figure3}a together with the relevant experimental data. An excellent agreement between data and computations is readily apparent both for the initial stage of defect closure and the post-yielding response.
\subsection{Fracture test}
The second step of the parameter calibration involves the parameters that control the response to tensile loading of the facets, thus leading to meso-scale fracture. In order to calibrate the meso-scale tensile strength, $\sigma_t$ and the tensile characteristic length, $\ell_t$, a direct tension test was simulated. In absence of specific information relevant to Bleurswiller sandstone, the area beneath the stress-displacement curve and the peak stress (Figure \ref{figure3}b) were compared with typical fracture energy and macroscopic tensile strength for sandstones found in the literature, 15 to 54 $\text{J/m}^\text{2}$ \cite{Atkinson_1987} and 0.3 to 8 MPa \cite{Hoagland_1972}, respectively. By setting $\sigma_t=2.5$ MPa, $\ell_t$=100 mm and $n_t$=0.1, provides $G_t$=30 $\text{J/m}^\text{2}$ and one can obtain a macroscopic fracture energy of $G_f$=33 $\text{J/m}^\text{2}$ and a macroscopic tensile strength of 3 MPa. It should be noted that macroscopic values of fracture energy is greater than the meso-scale value due to the presence of shear stresses in addition to the normal stresses on the facets which leads to the combination of tensile and frictional behaviors (mixed fracture mode) even under macroscopic mode 1 fracture conditions. The macroscopic tensile strength is also greater than the meso-scale tensile strength because the macroscopic peak stress is attained after stable crack propagation and local re-distribution at the meso-scale.

\subsection{Triaxial tests with low and high confining pressures}
To calibrate the model parameters controlling the shear behavior ($\sigma_s$=3.75 MPa, $\sigma_{N0}$=70 MPa, $\mu_0$=0.1 and $\mu_\infty$=0.05), two triaxial tests at different confining pressures were used, namely 10 MPa (to account for the brittle response typical of low confinement) and 100 MPa (to account for the response at high confinement). Figure \ref{figure3}c illustrates the response in terms of mean pressure, $p$, versus volumetric strain, while Figure \ref{figure3}d shows the differential stress, $q$, as a function of the axial strain.  It is possible to notice a good general agreement between data and LDPM computations, with the model being able to capture the brittle-ductile transition from low to high confinements, as well as the change from a dilative to a contractive volumetric response. While for high confinement the agreement is excellent from a quantitative standpoint, considerable differences can be noted between the amount of softening predicted by the model and that observed in the experiment at low confining pressure, with the model significantly underestimating the brittleness of the post-peak response. Such mismatch, however, can be explained as an outcome of the differences between the actual sample tested in the laboratory (the diameter of the rock cylinders was 40 mm and their length 80 mm) and that simulated by the LDPM (cylindrical specimens with the diameter and height of 2.4 mm). The role of this size-effect induced by damage localization and strain-softening, will be inspected numerically in the subsequent section.

\subsection{Response prediction for triaxial tests and size effect analysis}
\begin{figure}[!ht]
\centering
\includegraphics[width=7cm]{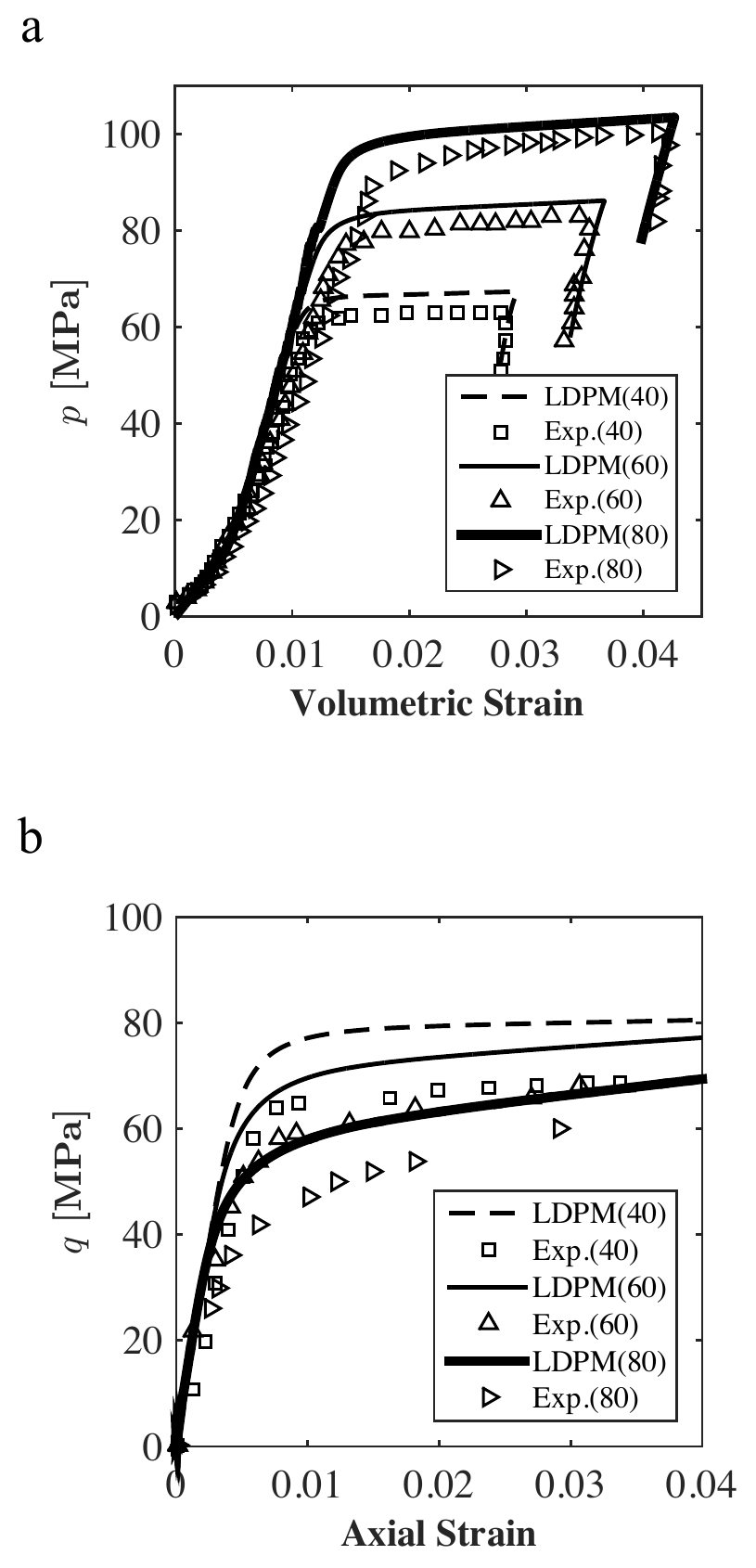}
\caption{(a) LDPM prediction of $p$ vs volumetric strain. (b) LDPM prediction of $q$ vs axial strain.}\label{figure4}
\end{figure}

\begin{figure}[!ht]
\centering
\includegraphics[width=7cm]{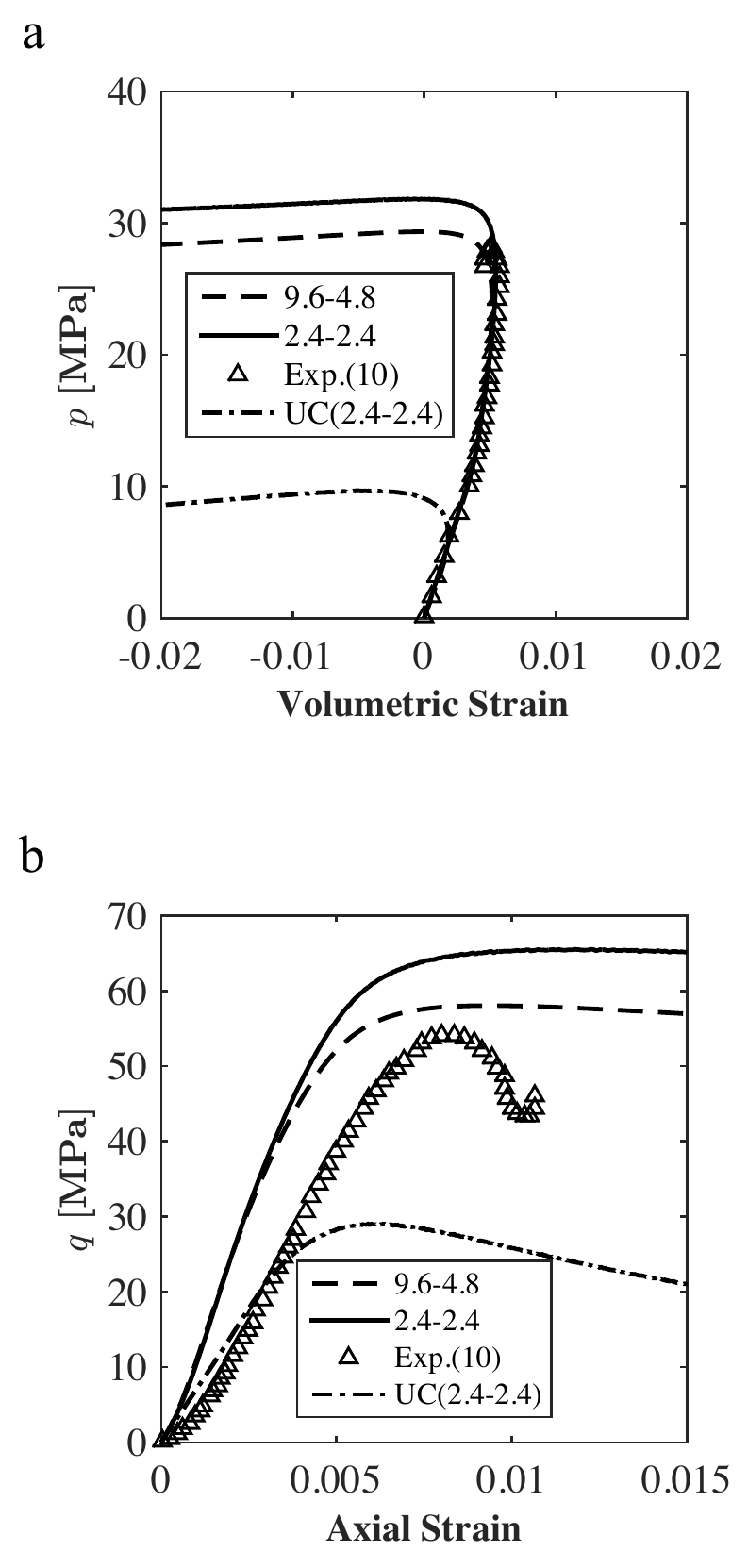}
\caption{Size effect study for low confinement triaxial test with 10 MPa confinement pressure and unconfined compression (UC) test.}\label{figure5}
\end{figure}

In this section, the calibrated LDPM is used to predict the response of compression tests performed at different confinement pressures (40, 60 and 80 MPa). The corresponding predictions are plotted in Figures \ref{figure4}a and \ref{figure4}b which demonstrate a good agreement between LDPM computations and experimental data, with LDPM capturing the pressure dependence of strength and compressibility, as well as the mean stress at the onset of shear-enhanced plastic compaction.\\
Additional triaxial tests were simulated to further explore the role of the size of the numerical sample in the brittle regime of deformation. This effect is illustrated in Figures \ref{figure5}a and \ref{figure5}b for 10 MPa confinement pressure for a cylindrical specimen with 9.6 mm height and 4.8 mm diameter with about 39,600 grains. It can be noticed that as the specimen size increases, the peak deviatoric stress predicted by the LDPM decreases and the amount of softening increases. This leads to an improvement of the model peformance in terms of post-peak behavior and dilative response, which tends to approach more closely the data. This result corroborates the constitutive choices made for the simulation of tensile fracturing at the meso-scale, indicating that the proposed model is capable of capturing the typical size dependence of the strength of quasi-brittle solids. As a result, although differences between data and simulations remain also in the case of a magnified numerical sample, it is arguable that such mismatch can be further mitigated by approaching the real size of the tested rock core.\\
In addition, the response for unconfined compression test demonstrates more softening and lower peak stresses in Figures \ref{figure5}a and \ref{figure5}b and consequently more brittleness of the mechanical behavior. It should be noted that the inelastic heterogeneity of the LDPM is the factor that automatically triggers and captures all different failure patterns using one set of meso-scale calibrated parameters and this feature gives superiority to the model compared to other existing discrete models.\\
For the unconfined compression test, the computed peak stress was 25 MPa corresponding to about 8 times the macroscopic tensile strength ($f_t$=3 MPa). This ratio is similar to typical values obtained for other quasi-brittle materials such as ceramics \cite{Grady_1994} and concrete  \cite{Peerlings_1998}, and therefore corroborates further the choices made for the selection of the  constitutive parameters.

%
\section{Comparison of LDPM results with macroscopic plasticity theories}\label{strain_localization analysis}
\begin{figure}[!ht]
\centering
\includegraphics[width=7cm]{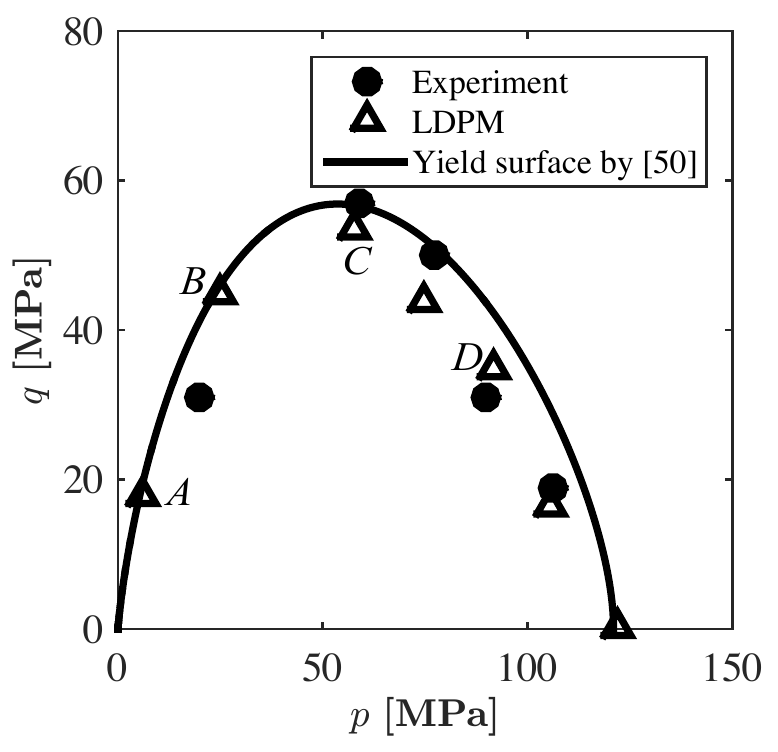}
\caption{Yield stresses of Bleurswiller sandstone measured through laboratory experiments (closed symbols), computed through LDPM (open symbols) and fitted through the yield surface expression proposed by \citet{Lagioia_1996}.}\label{figure6}
\end{figure}

\begin{figure}[!ht]
\centering
\includegraphics[width=7cm]{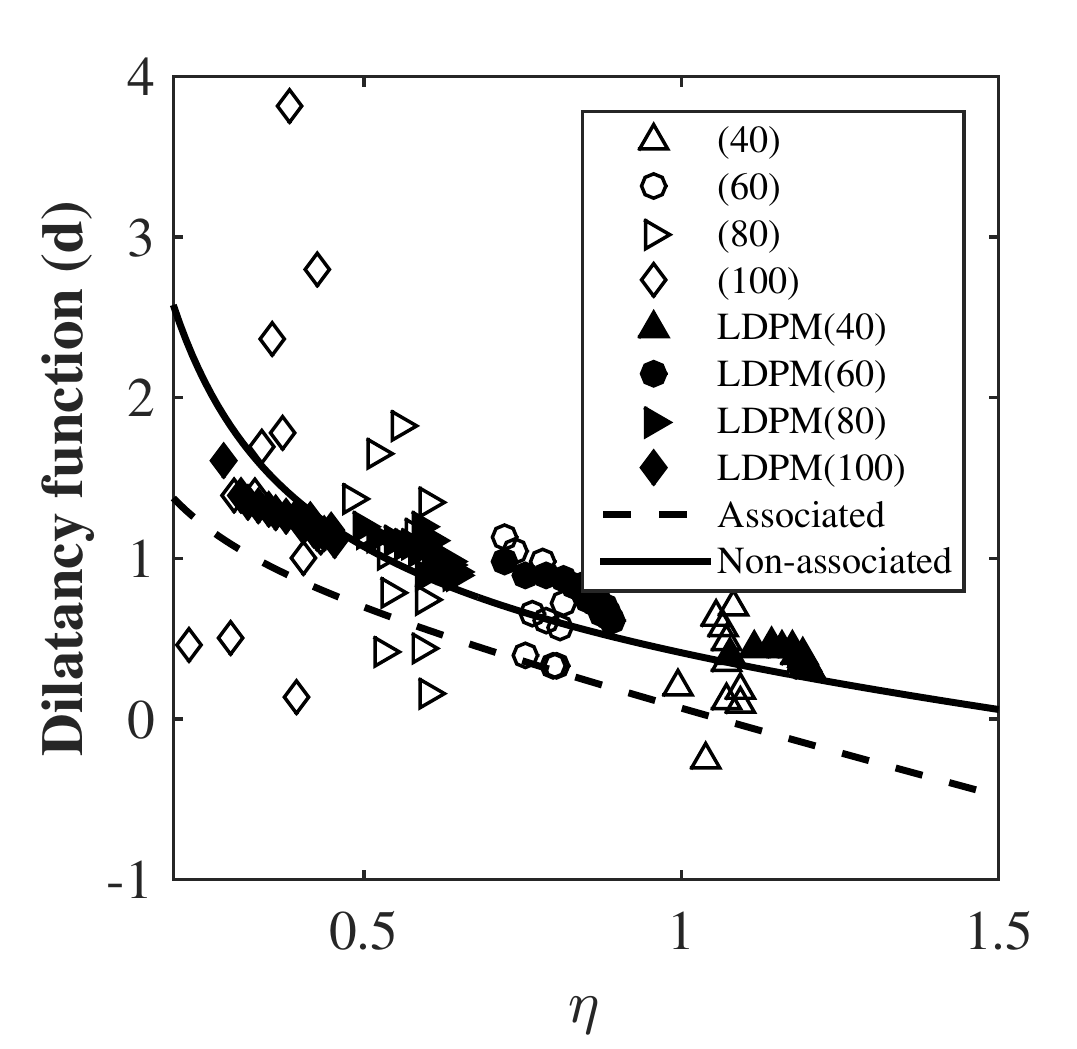}
\caption{Dilatancy function as a function of $\eta=q/p$.}\label{figure7}
\end{figure}

\begin{figure}[!ht]
\centering
\includegraphics[width=13cm]{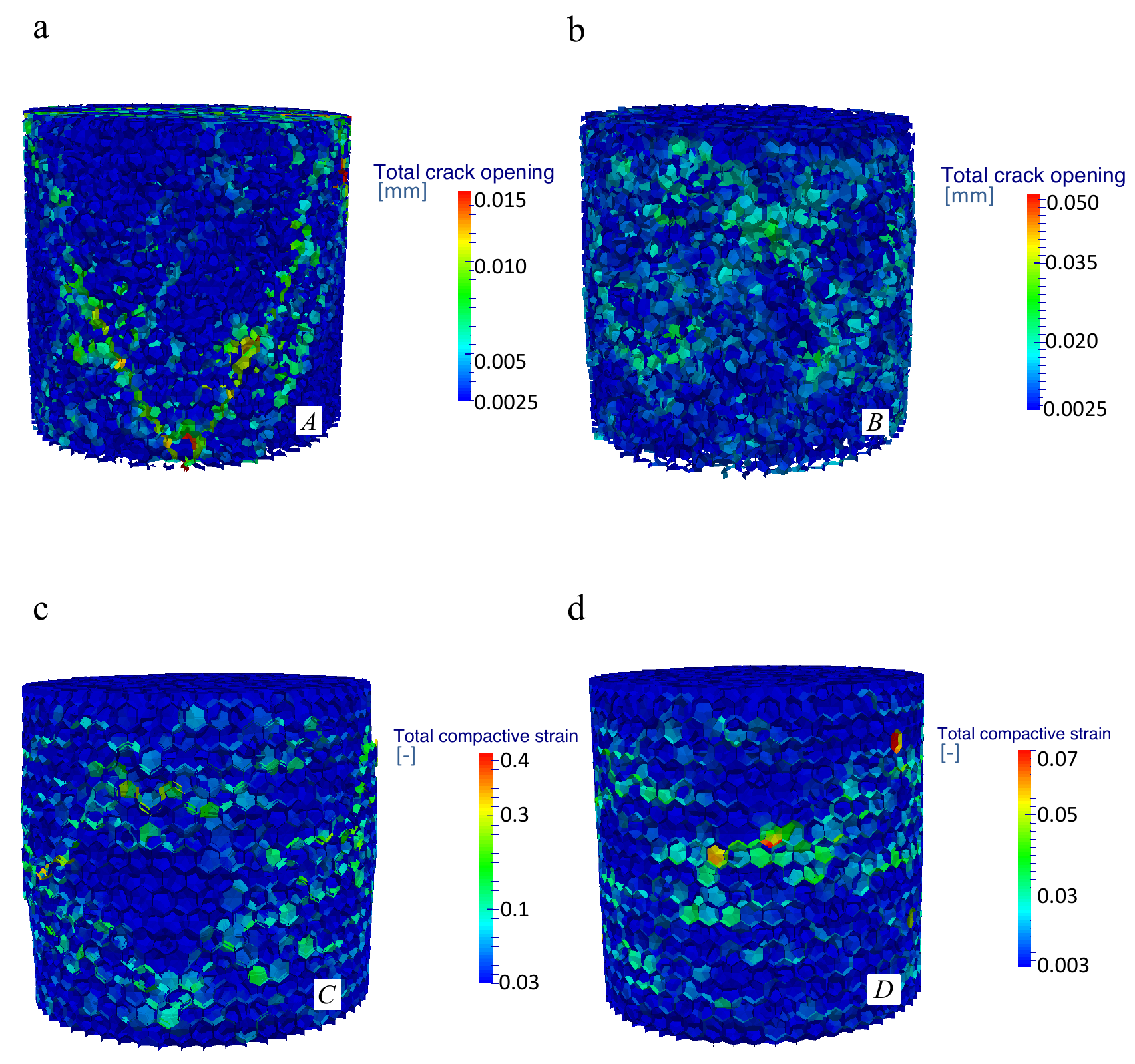}
\caption{(a),(b) LDPM fracture patterns for unconfined compression test and triaxial test with 10 MPa confinement pressure. (c),(d) LDPM compactive strain contours for triaxial tests with 40 MPa and 80 MPa confinement pressures.}\label{figure8}
\end{figure}

The previous sections have illustrated the ability of LDPM to simulate the pressure-dependent behavior of sandstones across the brittle and ductile regimes of deformation. To benefit of this capability, here LDPM is used as a virtual simulator to inspect classic concepts of rock plasticity, such as pressure-dependent yielding and plastic flow, as well as their impact on the strain localization characteristics.\\
Let us consider for this purpose the deformation response simulated for triaxial compression paths at varying levels of confinement. Each of the simulated stress-strain curves can be inspected to identify the points of deviation between linear and non-linear response. Such procedure identifies pressure-dependent yielding points, which can be plotted in the triaxial stress space, as customarily done for the interpretation of experiments (Figure \ref{figure6}). Considerable quantitative agreement can be noticed between data and LDPM predictions, with LDPM being capable of capturing the existence of a plastic cap at high-pressures as an emergent feature of the hypothesized meso-scale constitutive relations. This feature is consistent with continuum modeling techniques for porous rocks, as it is emphasized by the comparison between the LDPM-predicted yielding points and the shape of the yield surface proposed by \citet{Lagioia_1996}, which was recently used by Buscarnera and coworkers to simulate the plastic yielding of porous rocks of different mineralogy \cite{das_2014, Marinelli_2015}.\\
Another relevant comparison between the predictions of LDPM and the classical macroscopic description of rock inelasticity involves the predicted directions of plastic flow. A convenient strategy to explore the stress-dependence of this property involves the evaluation of the dilatancy function $d=\dot{e}_v^p/ \dot{e}_d^p$, i.e. the ratio between the increments of volumetric and deviatoric plastic strains. This function is plotted in Figure \ref{figure7} versus the stress ratio at yielding, $\eta=q/p$. Such plots are provided with reference to both experimental data (open symbols) and LDPM results (closed symbols). Despite the considerable scatter of the experimental data, an acceptable agreement can be observed, with LDPM simulations capable of capturing the decrease in volumetric flow components upon increasing values of stress ratio. Such trends can also be compared with the analytical expression between the dilatancy ratio and the stress ratio proposed by \citet{Lagioia_1996} (often referred to as stress-dilatancy relationship). Such functional relationship underpins a plastic potential compatible with the yield surface previously discussed with reference to Figure \ref{figure6}. In addition, it was recently used to study strain localization processes in porous rocks \cite{buscarnera_2014, Marinelli_2015}, and it can be readily used for the assessment of the degree of non-normality. The stress-dilatancy relationship proposed by \citet{Lagioia_1996} is characterized by the following expression:
\begin{equation}\label{etta_d}
d = \frac{\dot{e}^{p}_{v}}{\dot{e}^{p}_{d}} = \frac{\partial g/\partial p}{\partial g/\partial q} = \mu_{g}\left( M_{g} - \eta \right)
\left( \frac{\alpha_{g}M_{g}}{\eta} + 1 \right)\
\end{equation}
where $M_{g}$ represents the stress ratio at which plastic shearing takes place at constant volume (the so-called critical state), while $\alpha_{g}$ and $\mu_{g}$ are two shape parameters of the plastic flow rule. When Equation (\ref{etta_d}) is used in combination with the parameters that define the shape of the yield locus in Figure \ref{figure6} ($\mu_{g}=1.01, \alpha_{g}=0.11, M_{g}=1.06$; dashed line in Figure \ref{figure7}), the stress-dilatancy relationship provides a graphical representation of the plastic flow directions that would be predicted by a plasticity model based on an associated flow rule (i.e., it reflects the values of dilatancy ratio that would be produced by plastic flow directions oriented orthogonally to the yield surface reported in Figure \ref{figure6}). By contrast, if the same relation is adjusted to encompass the values of dilatancy ratio emerging from the data and/or the LDPM computations (solid line in Figure \ref{figure7}), a different set of parameters is obtained ($\mu_{g}=0.4, \alpha_{g}=0.45, M_{g}=1.6$). This result emphasizes the ability of LDPM to capture the macroscopic notion of non-associated plastic flow, which is here shown to guarantee a better fit of experimental data. In addition, this finding is compatible with classic bifurcation theories for plastic solids, according to which non-associativity is a key component to predict accurately the strain localization potential of cohesive-frictional materials  \cite{rudnicki_1975, issen_2000}.\\
To further validate the implications of the predicted non-normality of Bleurswiller sandstone, it is convenient to test the ability of LDPM to simulate the onset of pressure-dependent strain-localization. For this purpose, the predicted fracture patterns in terms of meso-scale total crack openings $w_{total}$ defined by Equation (\ref{total_crack_opening}) for facets with tensile normal strain, and compression bands in terms of compactive strains $e_{comp}$ defined by Equation (\ref{compactive_strain}) for facets with compressive normal strain can be used to map the spatial distribution of concentrated inelastic processes, and hence to identify the active zones of strain localization. One can write:
\begin{equation}\label{total_crack_opening}
w_{total} = \ell \sqrt{e_N^2+e_L^2+e_M^2 } \;\;\;(\text{for}\;\;e_N<0)
\end{equation}
and
\begin{equation}\label{compactive_strain}
e_{comp} = \sqrt{e_N^2+e_L^2+e_M^2 }  \;\;\;(\text{for}\;\;e_N>0)
\end{equation}
To study the effect of the mean pressure, four tests were chosen and their associated locations on the yield cap were marked in Figure \ref{figure6}: an unconfined compression test ($A$), and three triaxial tests at 10 MPa ($B$), 40 MPa ($C$) and 80 MPa ($D$) confinement. The total crack opening is selected as the metric for the interpretation of the simulations at the two lowest levels of confinement (brittle regime), while the total compactive strain is used for the interpretation of the two triaxial tests at the highest levels of confinement pressure.\\
Figure \ref{figure8}a depicts the fracture patterns computed for the unconfined compression test at $e_v=$-0.02 and $p=$7 MPa, illustrating the formation of  concentrated, inclined sub-vertical cracks typical of the brittle fracturing observed during unconfined compression. Figure \ref{figure8}b illustrates the results obtained for a triaxial test simulated at low confinement pressure of 10 MPa at $e_v=$-0.02 and $p=$31 MPa, thus displaying localized discrete fracture planes oriented along shear bands similar to those reported by \citet{fortin_2009} (shear-enhanced dilation and brittle faulting). For triaxial test simulated at 40 MPa confinement, \citet{fortin_2005} report a combination of localization bands oriented perpendicular to the maximum principal stress (compaction bands), and slightly inclined localization bands characterized by mixed shear/compaction deformation. Also in this case, both types of localized inelastic processes can be found in the numerical simulation illustrated in Figure \ref{figure8}c for $e_v=$0.03 and $p=$67 MPa. Finally, at the high confining pressure of 80 MPa at $e_v=$0.043 and $p=$103 MPa, the inelastic deformations predicted by LDPM are localized into several compaction bands nearly orthogonal to the maximum compressive stress, corresponding well to the deformation patterns reported by \citet{fortin_2005} at the same confinement pressure (Figure \ref{figure8}d).\\
\section{Conclusion}\label{conclusion}
Granular rocks exhibit pressure-dependent properties, as well as a broad range of strain-localization modes. Such materials are in fact characterized by various types of micro-scale heterogeneity, which generate macroscopic patterns that can be traced back to processes such as crack initiation; crack propagation; and interaction between fractured and unfractured material. Advanced multi-scale computations are thus required to simulate such patterns and correlate them with basic micro-scale processes. This paper has shown that LDPM is a framework able to fulfill such objectives for the important case of granular rocks. This feature has been discussed by presenting a strategy to incorporate into model computations grain-scale rock heterogeneity, i.e. the scale at which microscopic inelastic processes take place. A particular granular rock has been selected for model illustration purposes, the Bleurswiller sandstone, thus benefiting from the large availability of data about its mechanical response. The presented results show that by incorporating specific features, such as the crack closure upon compression and the development of pore collapse upon high-pressure compression, it is possible to capture a variety of macroscopic processes, such as the inelastic hydrostatic compression of rock samples, the brittle fracture upon tension, and the transition from
brittle/dilative response to ductile/compactive behavior. Most notably, the ability to predict such wide range of responses is based only on a limited set of data used for parametric identification, thus indicating that LDPM represents a versatile tool for a variety of geomechanical modeling applications, ranging from the intetpretation of multi-scale experiments, to the prediction of strain heterogeneities, to the formulation of continuum models and the assessment of their predictive capabilities.
\section{Acknowledgment} 
The authors would like to acknowledge the Institute for Sustainability and Energy at Northwestern (ISEN) funding scheme.
%
\section{References}
\bibliography{Shiva_rock}
%
\end{document}